\documentclass[10pt,journal,twocolumn]{IEEEtran}
\newif\ifCLASSOPTIONromanappendices \CLASSOPTIONromanappendicestrue
\usepackage{tikz}
\usepackage[siunitx]{circuitikz}
\usepackage[utf8]{inputenc}
\usepackage{fontenc}
\usepackage{nicematrix}
\usepackage{a0size}         
\usepackage{amssymb}
\usepackage{amsmath}
\usepackage{empheq}
\usepackage{upgreek}
\usepackage{arydshln}
\usepackage{multicol}
\usepackage[english]{babel}     
\usepackage{epsfig}
\usepackage{epstopdf}
\usepackage{bm}
\usepackage{changepage}
\usepackage{hyperref}
\usepackage{oplotsymbl}
\usepackage{bbm}
\usepackage{enumitem}
\usepackage{amsfonts,color,amsthm,amsmath, lscape,graphics}
\usepackage{subfiles}
\usepackage{comment}
\usepackage{algorithm} 
\usepackage{algpseudocode}

\usepackage[font=footnotesize]{caption}
\usepackage[font=footnotesize]{subcaption}

\usepackage[labelformat=simple]{subcaption}
\usepackage{algorithm} 
\usepackage{algpseudocode}
\usepackage{cite}
\usepackage{booktabs}

\def\imagunit{\mathsf{j}} % Imaginary number
\newcommand{\B}[1]{\boldsymbol{#1}}

\definecolor{awesome}{rgb}{1.0, 0.13, 0.32}
\usepackage{fix2col}

\addto\captionsenglish{}

\usepackage{graphicx}  
\theoremstyle{plain}

\usepackage{multirow} 
\usepackage{relsize}
\usepackage{textcomp, mathtools}
\usepackage{bm}   
\usepackage{pifont}
\usepackage{etoolbox}

\makeatletter
\patchcmd{\@maketitle}
  {\addvspace{0.5\baselineskip}\egroup}
  {\addvspace{-1\baselineskip}\egroup}
  {}
  {}
\makeatother
\begin{document}

\title{Guided Wireless Technology for \\ Near-Field Communication
\thanks{This work was supported by the Discovery Grants Program of the Natural Sciences and Engineering Research Council of Canada (NSERC) and the startup funding of the University of Tennessee.}
}
\author{\IEEEauthorblockN{Mohamed Akrout$^1$, Amine Mezghani$^2$, Faouzi Bellili$^2$, Robert W. Heath$^3$}\\
\IEEEauthorblockA{$^1$\textit{EECS Department}, \textit{University of Tennessee}, Knoxville, TN, USA, \\ $^2$\textit{ECE Department}, \textit{University of Manitoba}, Winnipeg, Canada \\ $^3$\textit{ECE Department}, \textit{University of California San Diego}, San Diego, CA, USA, \\
makrout@utk.edu, \{amine.mezghani,\,faouzi.bellili\}@umanitoba.ca, rwheathjr@ucsd.edu}
}

\maketitle

%%%%%%%%%%%%%%%%%%%%%%%%%%%%%%%%%%%%
%% Abstract
%%%%%%%%%%%%%%%%%%%%%%%%%%%%%%%%%%%%

\begin{abstract}
Guided wireless technology is an innovative approach that combines the strengths of guided waves and wireless communication. In traditional wireless systems, signals propagate through the air, where they are vulnerable to interference, attenuation, and jamming. Guided communication, in contrast, confines signals within a physical medium, significantly reducing interference and supporting higher data rates over longer distances. Guided wireless technology harnesses these benefits by creating guided wireless channels and offering a controlled pathway for electromagnetic waves. This work harnesses these benefits by focusing on the modeling of near-field communication through long connected arrays deployed in linear-cell environments. We then derive a circuit model for long array as an infinitely long dipole with multiple periodic feed points before approximating it with a finite array through open circuiting. Through our simulations, we show how the standing wave phenomenon is confirmed by the oscillations in spectral efficiency. We also demonstrate the capability of the LMMSE transmit beamformer in mitigating interference and minimizing the mean square error by adaptively allocating more power to the user experiencing the most severe channel attenuation, resulting in a more balanced variation of achievable rates across users.

%, and how the LMMSE transmit beamformer effectively mitigates interference by allocating power to the user experiencing the largest channel attenuation to minimize the mean square error, thereby ensuring a more balanced achievable rate between users.

%{\color{red} COMMENT: not sure about this lat statement, because usually LMMSE allocated more power to the weaker user (since we use the same common scalar gain at the receiver. Did you try with another location for user 1? maybe shift is by lambda/4, then we might get other oscillation behavior. )}
\end{abstract}

\begin{IEEEkeywords}
%\vspace{-0.18cm}
Reactive near-field, non-radiating communication, MIMO, circuit theory for communication, antenna arrays, linear cell environments.
\end{IEEEkeywords}

\section{Introduction}\label{Section 1}
\subsection{Background and Motivation}

Deployed communication solutions are increasingly integrating wired and wireless technologies to leverage the benefits of both while mitigating their drawbacks. Wired communication offers advantages such as vast, free bandwidth, minimal interference, low power operation, slowly time-varying channels, long-distance capabilities (especially with fiber optics), and high reliability and security. However, its deployment, particularly for the last mile, is challenging due to a significant environmental footprint, lack of flexibility, and poor scalability for a dynamic number of mobile users. This gap is filled by wireless communication, which provides mobility, reconfigurability, rapid deployment, and large coverage, but at the cost of significant interference and bandwidth issues due to free space radiation. In this work, we focus on wireless communication using reactive fields, i.e., communication in the near-field where the radiating system size $D$ is not significantly smaller than the distance $d$, that is $d = \mathcal{O}(D^{3/2})$. This is because reactive field-based communication, which offers several compelling advantages, including \textit{high security} due to the short communication range and dependence on precise device positioning, \textit{low environmental footprint} by mitigating wasteful radio emission and RF pollution, \textit{high energy efficiency} achieved through the confinement of EM interactions to the vicinity of devices, and \textit{low interference} as the specialized antenna arrays are less susceptible to external noise. While a common perception holds that reactive near-field communication is limited in range due to rapid signal attenuation, the variety of linear-cell environments (e.g., corridors, tunnels, buildings) presents promising deployment opportunities for large linear apertures. Depending on the electrical length of the linear aperture, the reactive near-field regime can occur at both short and mid-range distances, indicating its versatile potential for deployment.
%The three main communication infrastructure types, as depicted in Fig. \ref{fig:reactive-infrastructure}, are \textit{wired communication infrastructure} (using cables for critical, reliable applications), \textit{wireless communication using radiative fields} (transmitting over the air via radiative electromagnetic (EM) fields suitable for long distances, $D \ll d$), and 
% Integrating wired and wireless solutions overcomes the scalability issues faced by current high-densification wireless deployments, resulting in a sustainable and robust telecommunication infrastructure.
%This paper focuses on 
%For instance, in urban design, large linear apertures are consistent with standards for tall buildings \cite{saroglou2017towards}. 
\begin{figure}[h!]
    \centering
    \vspace{-0.3cm}
    \includegraphics[scale=0.2]{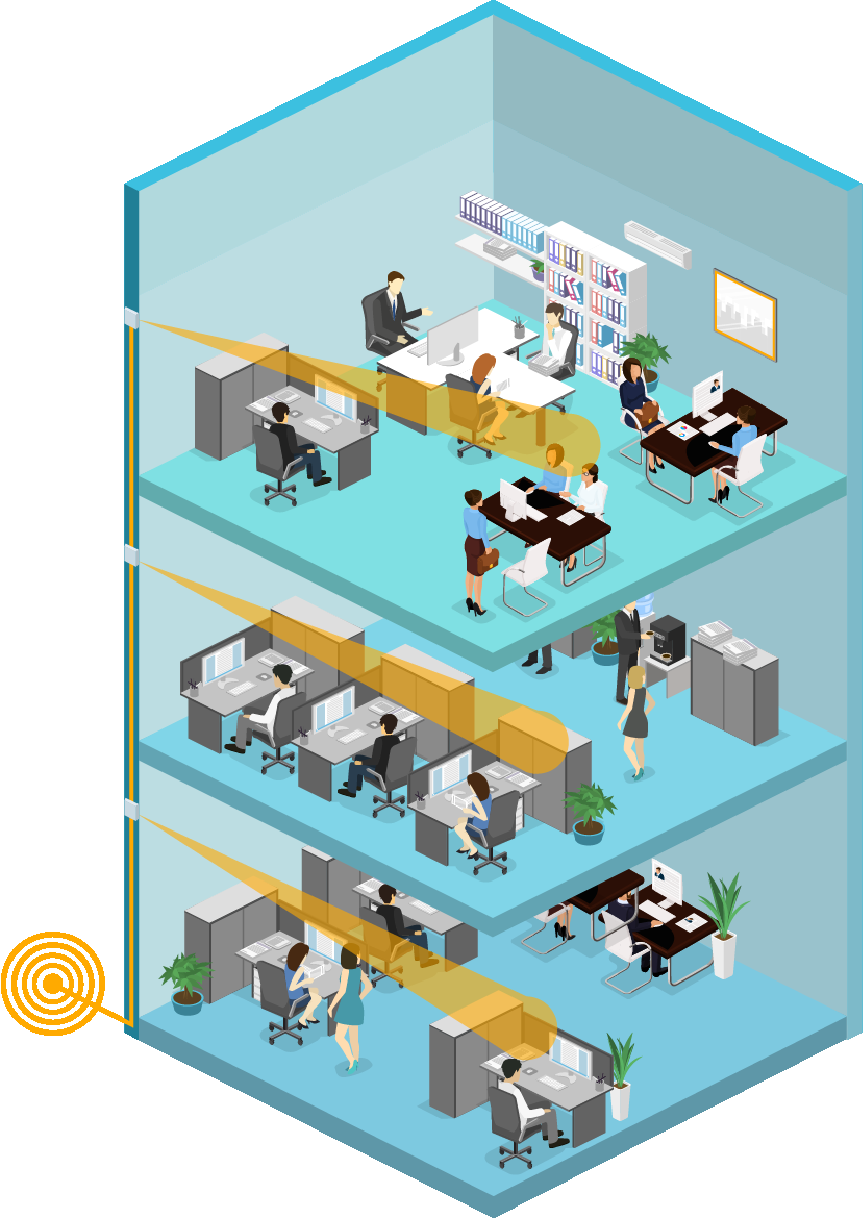}    
    \caption
    {wireless communication via reactive infrastructure (e.g., non-radiative antennas, wires, or cables)}.
    \label{fig:reactive-infrastructure}
    \vspace{-0.6cm}
\end{figure}

\subsection{Related Work}
The motivation behind reactive field-based communication is not new. One of the pioneering use cases is the ``G-Line'' transmission system \cite{GLine} in the mid-50s where bringing the TV signal to local communities was possible at ultra-high (UHF) frequencies using non-radiating wires instead of the traditional cable systems. As noted in \cite{GLine}, ``a home owner along the route of the G-Line simply placed his Yagi antenna close to the wire and got good TV reception''. Another important characteristic of these non-radiating wires is that they were unaffected by streets' noises (e.g., faulty electric motors, car ignitions), are simple to install and provide a uniform coverage. These non-radiating wires (a.k.a., leaky transmission lines or leaky feeders) were later  used for radiocommunication in underground mines and train tunnels \cite{martin1970radiocommunication,monk1956communication}, road vehicle communication \cite{nakamura1996road}, indoor cellular environments \cite{hou2019capacity}, guided radar \cite{gale1980comparative}, and wireless indoor positioning systems \cite{nakamura2010development}. We refer the reader to \cite{forooshani2013survey,martin1984leaky,hamid1977feasibility} for a comprehensive historical overview of non-radiating transmission lines. More recently, Ericsson has proposed radio stripes that are easy to deploy within future cell-free and distributed Massive MIMO \cite{interdonato2019ubiquitous}. The goal of this technology is to integrate antenna elements inside a single cable, thereby improving antenna arrangements for massive MIMO \cite{patentstripe}. However, Ericsson's radio stripes communicate through the radiating modes of emitted waves and hence are not designed to better store reactive EM fields for communication. In other words, Ericsson's radiative radio stripes are not devised toward reactive field-based communication unlike our proposed non-radiating linear apertures.

\subsection{Contribution}
We develop a circuit model for distributed connected arrays to accurately analyze and characterize near-field communication systems, including those utilizing reactive fields. The admittance matrix is more straightforward to derive due to the boundary condition on the electric field, but the impedance matrix of the infinite array is more suitable for capturing finite-size effects than the admittance matrix, as it naturally incorporates open-circuit conditions at both edges. Therefore, we first derive the admittance of the infinite connected array in the spatial Fourier domain, where the boundary condition on the electric field becomes tractable. The continuous admittance is then sampled and inverted to determine the impedance of the connected array at all the discrete excitation points, which directly reflects the physical feeding mechanism of the wire. This detailed modeling forms the basis for studying a MIMO communication system using a multi-port network, where the transmit is in the electric mode using the wire and the receiver is in the magnetic mode (e.g., using a loop antenna) to efficiently couple with the circulating magnetic field generated by the excited connected array.

\section{Circuit Model for Distributed Connected Array}\label{sec:metaline-circuit-model}

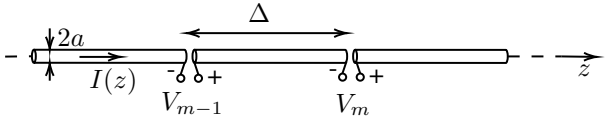
\begin{figure}[h!]
    \centering
    \vspace{-0.2cm}
    \tikzset{every picture/.style={line width=0.75pt}} %set default line width to 0.75pt        

\begin{tikzpicture}[x=0.75pt,y=0.75pt,yscale=-0.6,xscale=0.6]
%uncomment if require: \path (0,85); %set diagram left start at 0, and has height of 85

%uncomment if require: \path (0,85); %set diagram left start at 0, and has height of 85

%Shape: Can [id:dp004107362853040142] 
\draw   (96.66,33.97) -- (223.36,34.03) .. controls (224.27,34.03) and (225,36.49) .. (225,39.52) .. controls (225,42.55) and (224.26,45) .. (223.36,45) -- (96.65,44.93) .. controls (95.75,44.93) and (95.01,42.48) .. (95.01,39.45) .. controls (95.01,36.42) and (95.75,33.97) .. (96.66,33.97) .. controls (97.57,33.97) and (98.3,36.42) .. (98.3,39.45) .. controls (98.3,42.48) and (97.56,44.93) .. (96.65,44.93) ;
%Shape: Can [id:dp3189510221247136] 
\draw   (231.64,34) -- (358.35,34.07) .. controls (359.25,34.07) and (359.99,36.52) .. (359.99,39.55) .. controls (359.99,42.58) and (359.25,45.03) .. (358.34,45.03) -- (231.64,44.97) .. controls (230.73,44.97) and (230,42.51) .. (230,39.48) .. controls (230,36.45) and (230.74,34) .. (231.64,34) .. controls (232.55,34) and (233.29,36.46) .. (233.29,39.48) .. controls (233.29,42.51) and (232.55,44.97) .. (231.64,44.97) ;
%Shape: Can [id:dp6042306797792225] 
\draw   (366.65,34.03) -- (493.35,34.1) .. controls (494.26,34.1) and (494.99,36.56) .. (494.99,39.59) .. controls (494.99,42.61) and (494.25,45.07) .. (493.35,45.07) -- (366.64,45) .. controls (365.74,45) and (365,42.55) .. (365,39.52) .. controls (365,36.49) and (365.74,34.03) .. (366.65,34.03) .. controls (367.56,34.04) and (368.29,36.49) .. (368.29,39.52) .. controls (368.29,42.55) and (367.55,45) .. (366.64,45) ;
%Straight Lines [id:da41236589052783823] 
\draw    (135,40) -- (168,40) ;
\draw [shift={(170,40)}, rotate = 180] [color={rgb, 255:red, 0; green, 0; blue, 0 }  ][line width=0.75]    (10.93,-3.29) .. controls (6.95,-1.4) and (3.31,-0.3) .. (0,0) .. controls (3.31,0.3) and (6.95,1.4) .. (10.93,3.29)   ;
%Straight Lines [id:da7176261043099528] 
\draw    (539.99,39.59) -- (568,39.97) ;
\draw [shift={(570,40)}, rotate = 180.79] [color={rgb, 255:red, 0; green, 0; blue, 0 }  ][line width=0.75]    (10.93,-3.29) .. controls (6.95,-1.4) and (3.31,-0.3) .. (0,0) .. controls (3.31,0.3) and (6.95,1.4) .. (10.93,3.29)   ;
%Straight Lines [id:da6659066683647925] 
\draw    (223.36,45) -- (220,55) ;
%Straight Lines [id:da8613197365200453] 
\draw    (231.64,44.97) -- (235,55) ;
%Straight Lines [id:da10858756865640418] 
\draw    (359.36,44) -- (356,54) ;
%Straight Lines [id:da9699392895802417] 
\draw    (367.64,43.97) -- (371,54) ;
%Straight Lines [id:da7278624689996813] 
\draw  [dash pattern={on 4.5pt off 4.5pt}]  (494.99,39.59) -- (539.99,39.59) ;
%Straight Lines [id:da5352283497479102] 
\draw  [dash pattern={on 4.5pt off 4.5pt}]  (72.51,39.74) -- (95.01,39.45) ;
%Shape: Circle [id:dp4507279271963096] 
\draw   (353.17,57) .. controls (353.17,55.34) and (354.51,54) .. (356.17,54) .. controls (357.82,54) and (359.17,55.34) .. (359.17,57) .. controls (359.17,58.66) and (357.82,60) .. (356.17,60) .. controls (354.51,60) and (353.17,58.66) .. (353.17,57) -- cycle ;
%Shape: Circle [id:dp40450292024289825] 
\draw   (368,57) .. controls (368,55.34) and (369.34,54) .. (371,54) .. controls (372.66,54) and (374,55.34) .. (374,57) .. controls (374,58.66) and (372.66,60) .. (371,60) .. controls (369.34,60) and (368,58.66) .. (368,57) -- cycle ;
%Shape: Circle [id:dp7243144370280681] 
\draw   (217.17,57.98) .. controls (217.17,56.33) and (218.51,54.98) .. (220.17,54.98) .. controls (221.82,54.98) and (223.17,56.33) .. (223.17,57.98) .. controls (223.17,59.64) and (221.82,60.98) .. (220.17,60.98) .. controls (218.51,60.98) and (217.17,59.64) .. (217.17,57.98) -- cycle ;
%Shape: Circle [id:dp43072224244869894] 
\draw   (232,57.98) .. controls (232,56.33) and (233.34,54.98) .. (235,54.98) .. controls (236.66,54.98) and (238,56.33) .. (238,57.98) .. controls (238,59.64) and (236.66,60.98) .. (235,60.98) .. controls (233.34,60.98) and (232,59.64) .. (232,57.98) -- cycle ;
%Straight Lines [id:da38878450661877917] 
\draw    (110,16) -- (110,32) ;
\draw [shift={(110,34)}, rotate = 270] [color={rgb, 255:red, 0; green, 0; blue, 0 }  ][line width=0.75]    (10.93,-3.29) .. controls (6.95,-1.4) and (3.31,-0.3) .. (0,0) .. controls (3.31,0.3) and (6.95,1.4) .. (10.93,3.29)   ;
%Straight Lines [id:da9494384176413428] 
\draw    (110,60) -- (110,47) ;
\draw [shift={(110,45)}, rotate = 90] [color={rgb, 255:red, 0; green, 0; blue, 0 }  ][line width=0.75]    (10.93,-3.29) .. controls (6.95,-1.4) and (3.31,-0.3) .. (0,0) .. controls (3.31,0.3) and (6.95,1.4) .. (10.93,3.29)   ;
%Straight Lines [id:da5685208623352529] 
\draw    (227,20) -- (358,20) ;
\draw [shift={(360,20)}, rotate = 180] [color={rgb, 255:red, 0; green, 0; blue, 0 }  ][line width=0.75]    (10.93,-3.29) .. controls (6.95,-1.4) and (3.31,-0.3) .. (0,0) .. controls (3.31,0.3) and (6.95,1.4) .. (10.93,3.29)   ;
\draw [shift={(225,20)}, rotate = 0] [color={rgb, 255:red, 0; green, 0; blue, 0 }  ][line width=0.75]    (10.93,-3.29) .. controls (6.95,-1.4) and (3.31,-0.3) .. (0,0) .. controls (3.31,0.3) and (6.95,1.4) .. (10.93,3.29)   ;

% Text Node
\draw (206,49) node [anchor=north west][inner sep=0.75pt]   [align=left] {-};
% Text Node
\draw (240,49) node [anchor=north west][inner sep=0.75pt]   [align=left] {\mbox{+}};
% Text Node
\node[] at ($(220,60) + (10,20)$) {$\displaystyle V_{m-1}$};
% Text Node
\draw (342,48) node [anchor=north west][inner sep=0.75pt]   [align=left] {-};
% Text Node
\draw (376,48) node [anchor=north west][inner sep=0.75pt]   [align=left] {\mbox{+}};
% Text Node
\node[] at ($(355.17,60) + (10,20)$) {$\displaystyle V_{m}$};
% Text Node
\draw (141,47.4) node [anchor=north west][inner sep=0.75pt]    {$I(z)$};
% Text Node
\draw (551,44.4) node [anchor=north west][inner sep=0.75pt]    {$z$};
% Text Node
\draw (114,14.4) node [anchor=north west][inner sep=0.75pt]    {$2a$};
% Text Node
\node[] at ($(286,3.4) + (0,0)$)  {$\Delta $};

\end{tikzpicture}
    \vspace*{-0.8cm}
    \caption{Infinite uniform connected co-linear array.}
    \label{fig:colinear}
    \vspace*{-0.3cm}
\end{figure}

To model long wires composed of connected arrays, we consider the infinite co-linear one-dimensional array in Fig.~\ref{fig:colinear}, which can be treated as an infinitely long dipole with multiple feeds. The feed points are periodic and represent infinitesimal gaps distributed along the thin wire. By imposing the current as the primary boundary condition and then calculating the resulting voltages that develops across the gaps, the resulting impedance and open circuit field patterns provide the exact parameter needed to characterize the antenna load. To find the impedance of the distributed connected array, we first start by finding its admittance in the spatial Fourier domain because the boundary condition on the electric field makes it more practical to fix the voltages. We then use the inverse spatial Fourier transform to find the impedance of the array at all the discrete excitation points.

\subsection{Admittance of the connected array}
For a given wire having a magnetic constant $\mu$ and current density $\vec{J}(\vec{r},t)$ , we Maxwell's equation in potential form as follows 
\begin{equation}
    \left[\frac{1}{c^2} \frac{\partial^2}{\partial t^2}-\nabla^2 \right] \vec{A}(\vec{r},t)= \mu \,\vec{J}(\vec{r},t) 
\end{equation}
where the magnetic and electric fields, $\vec{B}(\vec{r},t)$ and $\vec{E}(\vec{r},t)$ are given by

\vspace{-0.2cm}
\begin{align}
 \vec{B}(\vec{r},t)   &=  \nabla \times \vec{A}(\vec{r},t) \label{B_field_} \\
   \frac{\partial \vec{E}(\vec{r},t) }{\partial t} &= c^2 
 \nabla \times  \vec{B} - c^2 \mu \vec{J}(\vec{r},t) \nonumber  \\
 &= c^2 \nabla (\nabla  \cdot \vec{A}(\vec{r},t))- \frac{\partial^2 \vec{A}(\vec{r},t) }{\partial t^2},\label{E_field_}
\end{align}
where $c$ is the speed of light. The general potential solution of (\ref{B_field_})--(\ref{E_field_}) is given by
\begin{equation}\label{eq:general-potential-sol}
 \vec{A}(\vec{r},t)= \frac{\mu}{4 \pi}    \iiint   \frac{\vec{J}(\vec{r},t-\frac{|\vec{r}-\vec{r}_0|}{c})}{|\vec{r}-\vec{r}_0|}{\rm d}V_0 
\end{equation}

\noindent Assuming a time harmonic solution for all quantities (i.e., time dependence proportional to ${\rm e}^{{\rm j} \omega t}$) which we will omit henceforth, the potential solution in (\ref{eq:general-potential-sol}) becomes:
\begin{equation}
 \vec{A}(\vec{r})= \frac{\mu}{4 \pi}    \iiint \vec{J}(\vec{r})  \frac{{\rm e}^{-{\rm j}k|\vec{r}-\vec{r}_0|}}{|\vec{r}-\vec{r}_0|}{\rm d}r 
 \label{solution_potential}
\end{equation} 

\begin{figure}[h!]
    \centering
    \includegraphics[scale=0.5]{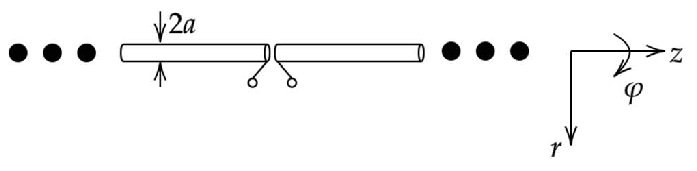}
    \caption{Cylindrical coordinates for an arbitrary wire portion of the connected dipole array.}
    \label{fig:cylindrical}
    \vspace{-0.1cm}
\end{figure}

To find the corresponding admittance matrix, we first examine the current distribution on a linear  cylindrical wire antenna center-driven by a delta function generator in the cylindrical coordinates $(r,\varphi,z)$. For a hollow-cylindrical wire antenna with a radius $a$ and infinite length as shown Fig. \ref{fig:cylindrical}, the vector potential in (\ref{solution_potential}) can be expressed in the cylindrical coordinate system based on with $\vec{J}(z,r)=\frac{I(z)}{2\pi a} \delta(r-a)\vec{\rm e}_z$ as
\begin{subequations}
\begin{align}
 & A_z(z, r )  =  \frac {\mu }{4\pi }  \int_{-\infty}^{\infty} I(\zeta)g(z-\zeta ) d \zeta, \label{Az_conv_} \\
& g(z,r )=  \frac {1}{2\pi } \int _ {-\pi }^ {\pi }  \frac {{\rm e}^ {-k R}}{R}  d  \varphi, \label{green_}  \\
& R= \sqrt{a^2+r^2-2 a r  \cos(\varphi) +z^ {2}}.
\end{align}
\end{subequations}
Here, $g(z,r)$ is the Green's function.

\noindent The boundary condition along the connected array ensures that the tangential electric field $E_z$ along the surface of the wire is equal to the voltage sources driving that wire at the center-driven feed points of each dipole. That is to say:
\begin{equation}
\begin{aligned}
      E_z  &\triangleq -  \frac{\imagunit}{\omega \mu\epsilon }  (  \partial^2 A_z(z,a)/   \partial{z}^{2}  +  k^{2}  A_z(z,a)) \\
      &= -\sum_{m=-\infty}^{\infty} V_m \delta(z-m\Delta),
\end{aligned}      
  \label{Ez_eq}     
\end{equation} 
where $\Delta$ is the uniform spacing between excitations along the infinite-long connected array, and $V_m$ is the voltage maintained at the $m$th feed point.

For ease of notation, we define the spatial Fourier transform of the Green's function $g(z,r)$ given in (\ref{green_}) in the wave number domain $\alpha$ as follows
\begin{equation}\label{eq:green-wavenumber}
G(\alpha,r )=\int\limits_{-\infty}^{\infty} g(\zeta,r  ) {\rm e}^{-{\rm j}\alpha \zeta} {\rm d}\zeta = -{\rm j} \pi  J_ {0}(\beta a)H_ {0}^{(2)}(\beta r  ),
\end{equation}
where $\beta= \sqrt{\kappa^2-\alpha^2}$, while $J_ {0}(\cdot)$ and $H_ {0}^{(2)}(\cdot )$ are the zero-order Bessel and Hankel functions. The equality in (\ref{eq:green-wavenumber}) is an established and fundamental result in the fields of classical electromagnetics, wave propagation, and antenna theory.

\noindent Next, after defining $\widehat{I}(\alpha)$ and $\widehat{V}(\alpha)$ as the current and voltage in the wave-number domain $\alpha$, we transform the convolution in (\ref{Az_conv_}) as:
\begin{equation}\label{eq:A-alpha}
    A(\alpha,r) = \frac {\mu }{4\pi } \, \widehat{I}(\alpha)\,G(\alpha, a).
\end{equation}

We also rewrite the boundary condition (\ref{Ez_eq}) in the $\alpha$ domain by injecting (\ref{eq:A-alpha})  as follows:
\begin{equation}
\frac{-\rm j}{\omega \mu \varepsilon}(k^2-\alpha^2) \frac{\mu}{4 \pi} \widehat{I}(\alpha)\,G(\alpha, a)=-\sum_{m=-\infty}^{\infty}V_m {\rm e}^{-{\rm j}\alpha m \Delta}=\widehat{V}(\alpha),
\end{equation}
from which we deduce the admittance in the Fourier domain as
\begin{equation}\label{eq:Y-alpha}
 Y(\alpha) =\frac{\widehat{I}(\alpha)}{\widehat{V}(\alpha)}=\frac {4k}{ Z_ {0}}  \frac{1}{\beta^{2}J_ {0}(\beta a)H_ {0}^{(2)}(\beta a )}.
\end{equation}

\noindent Finally, we use the inverse Fourier transform in the spatial domain to get the admittance impulse response as
\begin{equation}\label{current_dist_exact}  
\begin{aligned}
    y(z)&=\frac{1}{2\pi} \int_{-\infty}^{\infty}Y(\alpha)\,{\rm e}^{\imagunit \alpha z}\,d\alpha\\
    &= \frac {2k}{\pi Z_ {0}}  \int_{-\infty}^{\infty}  \frac{{\rm e}^{\imagunit \alpha z}}{\beta^{2}J_ {0}(\beta a)H_ {0}^{(2)}(\beta a )} d \alpha.
\end{aligned}
\end{equation} 

%An accurate approximation of the current distribution was found in \cite{Shen_1968}
%\begin{equation}\begin{aligned} & y(z)=\frac{I(z)}{V_0} \approx    \frac { \imagunit {\rm e}^ {\imagunit \kappa z}}{\zeta_{0}}     \\ & \cdot \log \left(  1-  \frac{2\pi \imagunit}{\log(\kappa z+\sqrt {(\kappa z)^ {2}+{\rm e}^ {-2\gamma }})-2\log(\kappa a)+\gamma +\imagunit\frac {\pi}{2}  }  \right)  \end{aligned}\label{current_dist_approx}  \end{equation}
%where $\gamma$ is the Euler's constant. This approximation is precise except for the imaginary part at $z=0$ (self-susceptance).

\subsection{Impedance of the connected array}
Until now, the spatial $z$ domain has been assumed continuous. To account for the discrete excitations along the distributed connected array and find the impedance at these feed points, let us assume a periodic excitation with uniform spacing $\Delta$ along the wire. In this case, the $n$th current excitation at $z_n=n\Delta$ not only due to the excitation at $z_n$ but also due to the sum of all the induced currents propagated along the wire from all the excitation points $\{z_m\}_{m=-\infty}^{+\infty}$. That is to say:
\begin{equation}\label{eq:I-z-wire-inf}
I(n \Delta)=\sum\limits_{m=-\infty}^{+\infty} V_m \, y(n\Delta-m\Delta).
\end{equation}

\noindent To maintain clarity between continuous-domain and discrete-domain representations, the subscript ``d'' will be used to denote discrete-time quantities. Using this notation, Ohm's law can be written at sampled spatial points after transforming (\ref{eq:I-z-wire-inf}) to the discrete space Fourier domain as follows:
\begin{equation}\label{eq:I-d}
    \begin{aligned}
        \widehat{V}(\alpha)\, Y_d(\alpha)&\triangleq \widehat{I}_d(\alpha)= \sum_{m=-\infty}^{+\infty}I(m \Delta ) \,{\rm e}^{-{\rm j}\alpha m \Delta}\\
        &= \frac{1}{\Delta}\sum\limits_{\ell=-\infty}^{\infty} \widehat{I}\Bigg(\alpha-\frac{2\pi \ell}{\Delta}\Bigg),
    \end{aligned}
\end{equation}
where

\begin{equation}\label{eq:Y-d}
Y_d(\alpha) = \frac{1}{\Delta}\sum\limits_{\ell=-\infty}^{\infty} Y(\alpha-\frac{2\pi \ell}{\Delta})=\frac{1}{Z_d(\alpha)}
\end{equation}

\noindent Both (\ref{eq:I-d}) and (\ref{eq:I-d}) follows from the the Poisson Summation Formula which states that the Fourier Transform of the sampled signal is equal to the periodic summation of the original signal's Fourier Transform.

Finally, the discrete impedance function in the spatial domain is obtained by transforming $1/Y_d(\alpha)$ using the inverse spacial Fourier transform as
\begin{equation}
\begin{aligned}
z_d[m \Delta]&=\frac{\Delta}{2 \pi} \int_0^{\frac{2\pi}{\Delta}}
  \frac{{\rm e}^{{\rm j}\alpha m \Delta }}{Y_d(\alpha) } {\rm d}\alpha \\
 & = \frac {Z_0\Delta^2}{8\pi \kappa}  \int_{0}^{\frac{2\pi}{\Delta}} {\frac{e^{\imagunit \alpha m\Delta}}{\sum\limits_{\ell=-\infty}^{\infty}\frac{1}{\beta_\ell^{2}J_ {0}(\beta_\ell a)H_ {0}^{(2)}(\beta_\ell a )}}} d \alpha,
 \end{aligned}
 \end{equation}
where $\beta_\ell=\sqrt{k^2-\alpha_{\ell}^2}$ is the propagation constant of the $\ell$th mode $\alpha_\ell$.

%\textcolor{red}{We derive the admittance matrix because the boundary condition on the electric field makes it more practical to fix the voltages. While our model assumes an infinite array, we only utilize a finite number of ports in practice, setting the others to open circuits to approximate the finite array case.}

\subsection{Fourier-Domain Field Pattern and Near-Field Coupling}
\noindent Defining $\beta=\sqrt{k^2-\alpha^2}$ and using (\ref{eq:Y-alpha}), we rewrite (\ref{eq:A-alpha}) as
\begin{equation}
\begin{aligned}
    A_z(\alpha,r)&=\frac{\mu}{4 \pi} \widehat{I}(\alpha)G(\alpha ,r )= \frac{\omega\mu \epsilon \widehat{V}(\alpha)}{{\rm j}\beta^2} \frac{H_0^{(2)}(\beta r)}{H_0^{(2)}(\beta a)}.
\end{aligned}
\end{equation}

\noindent The resulting magnetic potential in the cylindrical coordinate system then reads as 
\begin{equation}
H_\phi(r,\alpha)=-\frac{1}{\mu}\frac{\partial \widehat{A}_z(\alpha,r)}{\partial r}= \frac{\omega \epsilon \widehat{V}(\alpha)}{{\rm j}\beta^2} \frac{H_1^{(2)}(\beta r)}{H_0^{(2)}(\beta a)}.
\end{equation}

Considering a multi-excitation ports of the connected co-linear array, the current drives a dominant electric mode which generates a magnetic field that circulates concentrically around the array. To maximize signal reception, the receiver must couple efficiently with this field geometry. Employing a loop antenna (i.e., a magnetic dipole) is highly advantageous over a simple electric antenna because it strongly couples with the time-varying magnetic flux. By ensuring that the area of the loop antenna are receptive to the circulating magnetic field lines, the magnetic flux passing through the loop's area is maximized according to Faraday's law of induction. This will directly maximize the induced voltage and hence the received signal. For this reason, the receiver is designed to operate in the magnetic mode (e.g., using a loop antenna) to complement the electric mode operation of the wire transmitter.%, thereby enhancing modal coupling and maximizing the system's power transfer efficiency

 Assuming a magnetic Hertzian dipole as a probe antenna with magnetic current density $\vec{J}_{\rm M}(\B{r})=V' \delta\ell \delta(\B{r}-\B{r}_D)\vec{\rm e}_\phi $, the hybrid mutual coupling coefficient is given by
\begin{equation}
 \hat{h}_{12}^{\rm Hertz}(\alpha)=  \frac{V'\delta\ell \cdot H_\phi(r_D,\alpha)}{V' \hat{I}(\alpha) }= \frac{\delta  \ell\omega \epsilon Z_d(\alpha)}{{\rm j}\beta^2} \frac{H_1^{(2)}(\beta r)}{H_0^{(2)}(\beta a)}.
\end{equation}

For a magnetic Chu antenna with same radiation pattern as the magnetic Hertzian dipole outside a sphere of radius $a_{\rm Chu}$, we get the current coupling response based on the equivalence theorem established in \cite{akrout2022achievable} as
\begin{equation}
 \hat{h}_{12}^{\rm Chu}(\alpha)= \frac{\hat{I}^{\rm Chu}(\alpha)}{\hat{I}_d(\alpha )}=\frac{\sqrt{{\rm real}\{Y^{\rm Chu}\}}}{\sqrt{{\rm real}\{Y^{\rm Hertz}\}}} \hat{h}_{12}^{\rm Hertz}(\alpha),
 \end{equation}
 with ${\rm real}\{Y^{\rm Hertz}\}=\frac{2\pi \delta \ell^2}{3Z_0 \lambda^2}$.

 \section{Multi-antenna/Multi-User  Circuit Model}\label{sec:circuit-mimo-model}
 
%\subsection{Transmit and receive mode considerations}
%The connected co-linear array transmits a signal by generating a dominant electric mode, which results in a concentric circulating magnetic field. To achieve maximum signal reception, the receiver must exploit the magnetic field geometry. Therefore, a magnetic dipole (loop antenna) is chosen because its design allows it to strongly couple with the time-varying magnetic flux, maximizing the induced voltage via Faraday's Law. This magnetic mode reception successfully complements the electric mode transmission of the wire, ensuring high efficiency.

%For this reason, we allow the receiver to operate in the magnetic mode while the transmitter is operating in the electric mode.%This superior geometric coupling and modal alignment afford two key benefits: a more effective path for power exchange, and enhanced immunity to local electric field noise compared to an electric antenna, thereby maximizing the overall Signal-to-Noise Ratio (SNR).
%\textcolor{red}{why transmitter must be electric and why receiver must be magnetic}

\begin{figure}[t!]
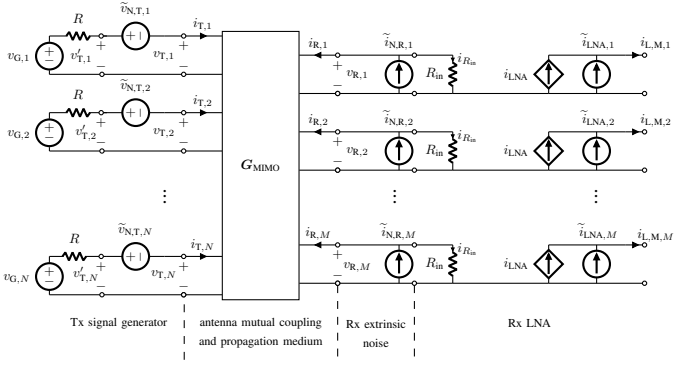

\centering
  \subfile{./figs/magelec-MIMO-communication-circuit-model_different_M_N_asilomar.tex}
  \vspace{-0.2cm}
  %\subfile{./figures/MIMO-communication-circuit-model.tex}
\caption{Linear multiport model of a radio MIMO communication system showing signal generation, antenna mutual coupling, and noise from both extrinsic (i.e., picked up by the antennas) and intrinsic (i.e., generated by LNAs and local circuitry) origins.}
\label{fig:MIMO-communication-circuit-elecmag}
\vspace{-0.2cm}
\end{figure}

%\subsection{Circuit model for MIMO communication}
We consider the circuit model depicted in Fig. \ref{fig:MIMO-communication-circuit-elecmag} to study a MIMO communication system with a transmitter having $N$ electrical-mode excitations and $M$ magnetic-mode receive antennas, each associated to a single-antenna user. There, the joint matrix $\bm{G}_\text{MIMO}$ models the overall circuit matrix accounting for the wireless channel and the mutual coupling effects. Specifically, we write Ohm's law for the multiport network The $N$ transmit voltage sources are represented by non-ideal voltage generators $\bm{v}_{\textrm{G}}(f)=\big[ v_{\text{G},1}(f), v_{\text{G},2}(f), \dots, v_{\text{G},N}(f)\big]^\top$, i.e., with internal resistance $R$. Their terminals are connected to the transmitting antennas with the current-voltage pairs $(\bm{v}_\textrm{T}(f),\bm{i}_\textrm{T}(f))$ through the noise voltage sources $\widetilde{\bm{v}}_{\textrm{N,T}}(f)= \big[ \widetilde{v}_{\text{N,T},1}(f), \widetilde{v}_{\text{N,T},2}(f), \dots, \widetilde{v}_{\text{N,T},N}(f)\big]^\top$. Likewise, the receive antenna terminals with the current-voltage pair $(\bm{v}_\textrm{R}(f),\bm{i}_\textrm{R}(f))$ are connected to the low-noise amplifiers (LNAs) through the noise voltage sources $\widetilde{\bm{v}}_{\textrm{N,R}}(f)= \big[ \widetilde{v}_{\text{N,R},1}(f), \widetilde{v}_{\text{N,R},2}(f), \dots, \widetilde{v}_{\text{N,R},M}(f)\big]^\top$. These LNAs are modeled as noisy frequency flat devices with gain $\beta$. Finally, the LNA is connected to the outside world through the output current port $\bm{i}_\textrm{L}(f) = \big[ i_{\text{L},1}(f), i_{\text{L},2}(f), \dots, i_{\text{L},M}(f)\big]^\top$ to which a load impedance of a device can be connected. The relationship between the transmit/receive
voltages and currents of $\bm{G}_\text{MIMO}$ is given by Ohm's law as \vspace{-0.1cm}
%{\color{red} COMMENT: we should remove the subscript R and M, right, as defined above? Same for the appendix}
%Since the circuit model in Fig.~\ref{fig:MIMO-communication-circuit} accounts for the extrinsic noise in the transmit/receive antennas and LNAs for more realistic scenarios, it is necessary to specify the statistical properties of all noise voltage sources as we describe hereafter.
%\subsubsection{Multiport description}:
\begin{equation}\label{eq:initial-circuit-equation}
    \left[\begin{array}{l}
\bm{v}_{\textrm{T}} \\
\bm{i}_{\textrm{R}}
\end{array}\right]~=~ \underbrace{\left[\begin{array}{cc}
\bm{Z}_{\textrm{T}} & \bm{H}_{\textrm{TR}}\\
\bm{H}_{\textrm{RT}} & \bm{Y}_{\textrm{R}}
\end{array}\right]}_{\triangleq~\bm{G}_{\textrm{MIMO}}}\,\left[\begin{array}{l}
\bm{i}_{\textrm{T}} \\
\bm{v}_{\textrm{R}}
\end{array}\right],
\vspace{-0.2cm}
\end{equation}
where $\bm{Z}_{\textrm{T}}$ and $\bm{Y}_{\textrm{R}}$ are the transmit impedance and receive admittance while $\bm{H}_{\textrm{TR}}$ and $\bm{H}_{\textrm{RT}}$ are the hybrid matrices modeling the transmit-receive and receive-transmit coupling, respectively.

The relationship between the receive output current $\bm{i}_{\textrm{L},\textrm{M}}$ and the transmit voltage generators $\bm{v}_{\textrm{G}}$ can be obtained as follows (cf. Appendix \ref{appendix:inout}):
%Assuming the same properties of the noise voltages/currents as in \cite{ivrlavc2010toward, akrout2023super}, the input properties
\begin{equation}\label{eq:inout}
    \bm{i}_{\textrm{L},\textrm{M}}=\bm{H}\,\bm{v}_{\textrm{G}} + \bm{n},
\end{equation}
where
\begin{subequations}\label{eq:inout2}
    \begin{align}
        \bm{H} &=\frac{\beta}{R_{\textrm{in}}}\,\bm{K}_{\textrm{RT}},\\
        \bm{n}&= \frac{\beta}{R_{\textrm{in}}}\,\bm{K}_{\textrm{RT}}\,\bm{\widetilde{v}}_{\textrm{T},\textrm{E}} + \frac{\beta}{R_{\textrm{in}}}\,\bm{K}_{\textrm{R}}\,\bm{\widetilde{i}}_{\textrm{R},\textrm{E}} + \widetilde{\bm{i}}_{\textrm{LNA}},
    \end{align}
\end{subequations}
\noindent with the matrices $\bm{K}_{\textrm{RT}}$ and $\bm{K}_\textrm{R}$ being defined in (\ref{eq:Ks}). Assuming the same properties of the noise voltages/currents as in \cite{ivrlavc2010toward, akrout2023super}, it follows that the noise correlation matrix is given by:
\begin{subequations}\label{eq:noise-correlation}
\small{
\begin{align}
    \bm{R}_{\bm{n}} &= \frac{4\,k_b\,T\,\beta^2}{R_{\textrm{in}}^2}\, \Bigg[\frac{(N_{\textrm{f}}-1)\,R_{\textrm{in}}}{\beta}\,\mathbf{I}_M + \bm{K}_{\textrm{RT}}\,\Re{\{\bm{Z}_{\textrm{T}}\}}\,\bm{K}_{\textrm{RT}}^{\mathsf{H}} \nonumber \\
    &\hspace{2cm}+ ~\bm{K}_{\textrm{RT}}\,\Re{\{\bm{H}_{\textrm{TR}}\}}\,\bm{K}_{\textrm{R}}^{\mathsf{H}}+\bm{K}_{\textrm{R}}\,\Re{\{\bm{H}_{\textrm{RT}}\}}\,\bm{K}_{\textrm{RT}}^{\mathsf{H}} \nonumber\\
    &\hspace{2cm}+ \bm{K}_{\textrm{R}}\,\Re{\{\bm{Y}_{\textrm{R}}\}}\,\bm{K}_{\textrm{R}}^{\mathsf{H}}\Bigg].
\end{align}
}
\end{subequations}

\section{Numerical results and discussions}\label{sec:results}

In this section, we simulate a connected array using the impedance model developed in Section \ref{sec:metaline-circuit-model} and employ the communication system model in Section \ref{sec:circuit-mimo-model} to compute the spectral efficiency. It is important to note that our simulations are independent of the frequency because our connected array model depends on the distance $d$ through the factor $kd$.

\begin{figure}[h!]
    \centering
    \includegraphics[scale=0.2]{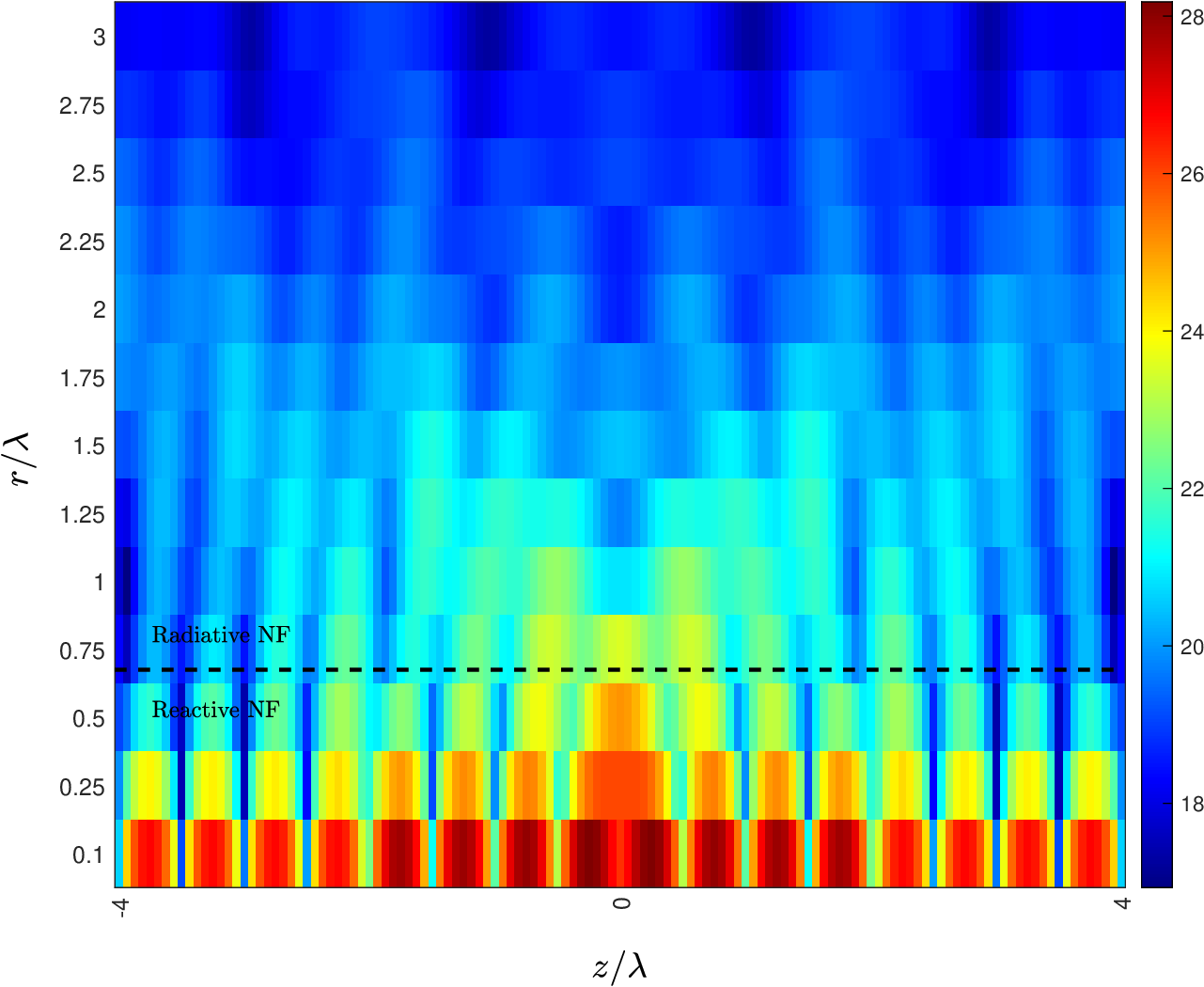}
    \caption{Spectral efficiency for spacing $\Delta = 2\,\lambda$.}
    \label{fig:rate-spacing-2lambda}
    \vspace{-0.3cm}
\end{figure}

\begin{figure}[h!]
    \centering
    \includegraphics[scale=0.2]{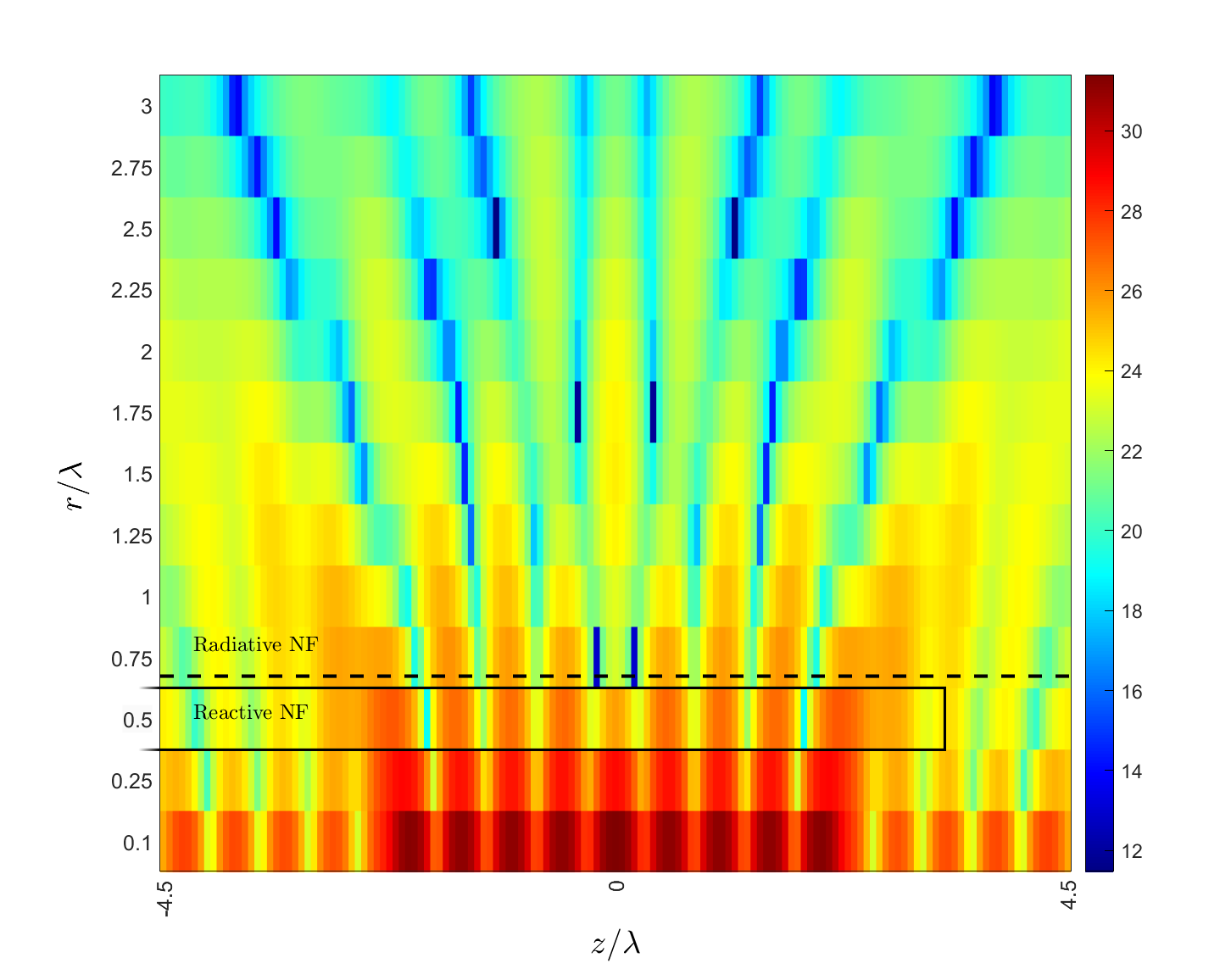}
    \caption{Spectral efficiency for spacing $\Delta = 2.25\,\lambda$.}
    \label{fig:rate-spacing-2.25lambda}
    \vspace{-0.1cm}
\end{figure}

In Fig. \ref{fig:rate-spacing-2lambda}, we plot the spectral efficiency of the connected array placed along the $z$-axis and allow a single excitation at $z=0$. We observe the standing waves as the scattering occurs between the connected dipoles for spacing $\Delta=2\lambda$. We plot the same thing in Fig. \ref{fig:rate-spacing-2.25lambda}, but for a slightly larger spacing $\Delta=2.25\lambda$, and we observe how we get more constructive waves around the connected array due to the constructive interference at this spacing.

%[improve: we simulate a connected array using impedance model developed in Section II and use the system model in Section III to compute the achievabe rate. Note that our simulations are independent of the frequency because our connected array model depends on the distance $d$ through the factor $kd$.].
%[improve: in fig 5 we plot the spectral efficiency of the connected array place along the z axis at r=0 and we allow a single excitation at z=0 we observe the standing waves as the scattering occurs between the connected dipoles for spacing $\Delta=2\,\lambda$, we plot the same thing but for spacing $\Delta=2.25\,\lambda$ in fig 6 and see how we get more constructive waves around the connected array because ...]

\begin{figure}[t]
    %\centering
    \begin{subfigure}[b]{0.2\textwidth}
        \includegraphics[scale=0.35]{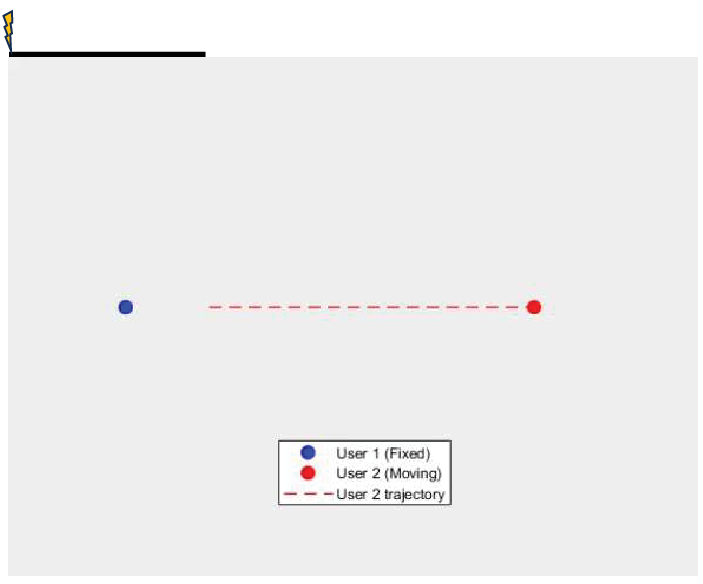}
        \vspace{0.15cm}
        \caption{}
        \label{fig:2user-comm}
    \end{subfigure}
    \hspace{0.3cm}
    \begin{subfigure}[b]{0.2\textwidth}
        \includegraphics[scale=0.23]{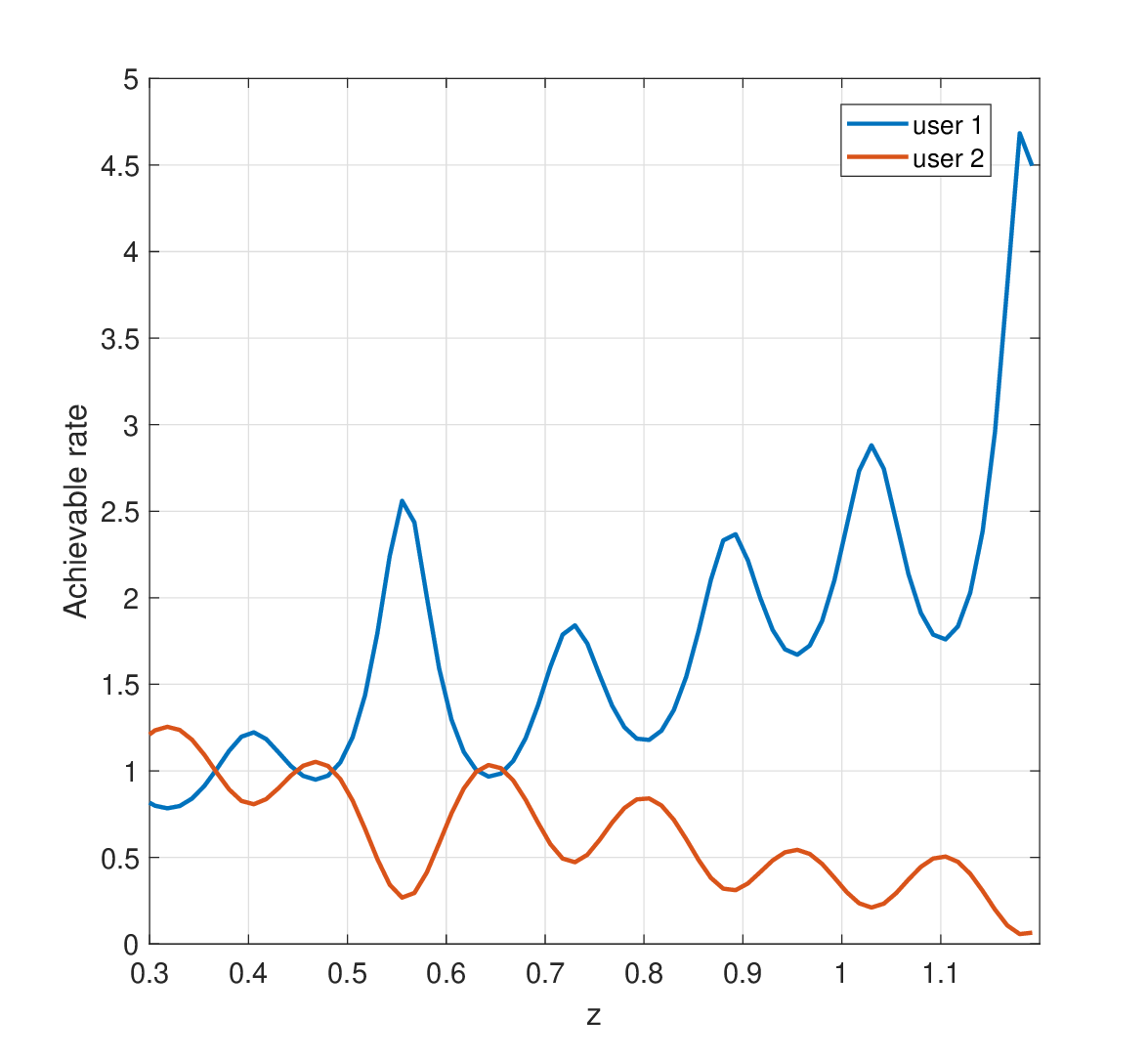}
        \caption{}
        \label{fig:2user-achievable-rate}
    \end{subfigure}
    \caption{(a) Two-user scenario where user 1 is fixed and user 2 is moving along a line, (b) achievable rate of the users over the z-axis.}
    \label{fig:2user-experiment}
\end{figure}

Finally, we consider a two-user scenario as depicted in Fig. \ref{fig:2user-comm}  where user 1 (in blue) is fixed while user 2 (in red) is moving along a horizontal line parallel to the axis of the connected array. Using the linear minimum mean square error (LMMSE) transmit beamformer \cite{joham2005linear}, we plot the achievable rate of the two users in Fig. \ref{fig:2user-achievable-rate}, and we observe how they oscillate, reflecting the standing wave nature at the reactive near-field. Furthermore, it is seen that the two plots have opposite peaks. This confirms the interference mitigation capability of the LMMSE transmit beamformer which adaptively allocates power to the user experiencing the greatest channel attenuation to minimize the mean square error, thereby ensuring a more balanced achievable rate between users. %This confirms the interference mitigation capability of the LMMSE beamformer, which adaptively allocates power to the user with the stronger effective channel, thereby compensating for the channel attenuation experienced by the second user {\color{red} see my comment in the abstract}.

%[improve: Finally, we consider a 2-user scenario as depicted in Fig.  \ref{fig:2user-comm} where user 1 (in blue) is fixed while user 2 (in red) is moving long a horizontal line parallel to the axis of the connected array. Using the LMMSE transmit beamformer \cite{joham2005linear}, we plot the achievable rate of the two users and we see how they oscillate reflecting the standing wave nature at the reactive near field, also we see that the two plots have opposite peaks when one at max, the other is at min confirming that the LMMSE is doing its job, what it a user at a channel dip, it focuses on the other user with better channel conditions.]

%\clearpage 
\begin{appendices}
\renewcommand{\thesectiondis}[2]{\Roman{section}:}
\section{Derivation of the input/output relationship of the MIMO circuit model}\label{appendix:inout}

Here, we derive the input-output relationship between the input voltage vector $\bm{v}_{\textrm{G},\textrm{E}}$ and the output current vector $\bm{i}_{\textrm{L},\textrm{M}}$. For the circuit representing the electric mode, we apply the Kirchhoff's voltage law (KVL) between the two closed loops on the left- and right-hand sides of the joint channel matrix $\bm{G}_{\textrm{MIMO}}$ in Fig. \ref{fig:MIMO-communication-circuit-elecmag}. By doing so, one obtains:
\begin{subequations}\label{eq:vTE-iRM-KVL-KCL}
    \begin{align}
    \bm{v}_{\textrm{T}} &= \bm{v}_{\textrm{G}} - R\,\bm{i}_{\textrm{T}} + \bm{\widetilde{v}}_{\textrm{N},\textrm{T}},\\
    \bm{i}_{\textrm{R}} &= \bm{\widetilde{i}}_{\textrm{N},\textrm{R}} - \frac{1}{R_{\textrm{in}}}\,\bm{v}_{\textrm{R}}.
    \end{align}
\end{subequations}

\noindent We also apply the KCL laws to the LNA circuit to obtain:
\begin{equation}\label{eq:iLM-vRM}
    \bm{i}_{\textrm{L},\textrm{M}} =\bm{\widetilde{i}}_{\textrm{LNA}} +\frac{\beta}{R_{\textrm{in}}}\,\bm{v}_{\textrm{R}}.
\end{equation}

\noindent After injecting (\ref{eq:vTE-iRM-KVL-KCL}) back into (\ref{eq:initial-circuit-equation}) and performing some algebraic manipulations, we get:
\begin{equation}\label{eq:initial-circuit-equation-2}
\underbrace{\left[\begin{array}{cc}
\bm{Z}_{\textrm{T}} + R\,\mathbf{I}_N & \bm{H}_{\textrm{TR}}\\
\bm{H}_{\textrm{RT}} & \bm{Y}_{\textrm{R}} + \frac{1}{R_{\textrm{in}}}\,\mathbf{I}_M
\end{array}\right]}_{\triangleq \, \bm{F}_{\textrm{MIMO}}}\,\left[\begin{array}{l}
\bm{i}_{\textrm{T}} \\
\bm{v}_{\textrm{R}}
\end{array}\right] = \left[\begin{array}{l}
\bm{v}_{\textrm{G}} \\
\bm{0}_M
\end{array}\right] + \left[\begin{array}{l}
\bm{\widetilde{v}}_{\textrm{N},\textrm{T}} \\
\,\bm{\widetilde{i}}_{\textrm{N},\textrm{R}}
\end{array}\right].
\end{equation}

\noindent For ease of notation, we define the following quantity:
\begin{equation}
    \bm{Q} = \left(\bm{Z}_{\textrm{T}} + R\,\mathbf{I}_N - \bm{K}_{\textrm{TR}} \left(\bm{Y}_{\textrm{R}} + \frac{1}{R_{\textrm{in}}}\,\mathbf{I}_M\right)^{-1} \bm{K}_{\textrm{RT}}  \right)^{-1}. 
\end{equation}

\noindent Let $\bm{F}_{\textrm{MIMO}}^{-1} \triangleq \bm{K}_{\textrm{MIMO}} = \left[\begin{array}{cc}
\bm{K}_{\textrm{T}} & \bm{K}_{\textrm{TR}}\\
\bm{K}_{\textrm{RT}} & \bm{K}_{\textrm{R}}
\end{array}\right] $ with the block matrices given by:
\begin{subequations}\label{eq:Ks}
    \begin{align}
        \bm{K}_{\textrm{T}} &= \bm{Q},\\
        \bm{K}_{\textrm{TR}} &= -\bm{Q} \, \bm{H}_{\textrm{TR}} \left(\bm{Y}_{\textrm{R}} + \frac{1}{R_{\textrm{in}}}\,\mathbf{I}_M\right)^{-1},\\
        \bm{K}_{\textrm{RT}} &= - \left(\bm{Y}_{\textrm{R}} + \frac{1}{R_{\textrm{in}}}\,\mathbf{I}_M\right)^{-1} \bm{H}_{\textrm{RT}}\,\bm{Q},\\
        \bm{K}_{\textrm{R}} &= \left(\bm{Y}_{\textrm{R}} + \frac{1}{R_{\textrm{in}}}\,\mathbf{I}_M\right)^{-1} + \left(\bm{Y}_{\textrm{R}} + \frac{1}{R_{\textrm{in}}}\,\mathbf{I}_M\right)^{-1} \\
        &\hspace{1cm}\times\bm{H}_{\textrm{RT}}\,\bm{Q}\,\bm{H}_{\textrm{TR}}\,\left(\bm{Y}_{\textrm{R}} + \frac{1}{R_{\textrm{in}}}\,\mathbf{I}_M\right)^{-1}.
    \end{align}
\end{subequations}

\noindent Multiplying the two sides of (\ref{eq:initial-circuit-equation-2}) by $\bm{K}_{\textrm{MIMO}}$ yields:
\begin{equation}\label{eq:initial-circuit-equation-3}
\begin{aligned}
    \left[\begin{array}{l}
\bm{i}_{\textrm{T}} \\
\bm{v}_{\textrm{R}}
\end{array}\right] = \left[\begin{array}{cc}
\bm{K}_{\textrm{T}} & \bm{K}_{\textrm{TR}}\\
\bm{K}_{\textrm{RT}} & \bm{K}_{\textrm{R}}
\end{array}\right]\left[\begin{array}{l}
\bm{v}_{\textrm{G}} \\
\bm{0}_M
\end{array}\right] \\
&\hspace{-3cm}+ \left[\begin{array}{cc}
\bm{K}_{\textrm{T}} & \bm{K}_{\textrm{TR}}\\
\bm{K}_{\textrm{RT}} & \bm{K}_{\textrm{R}}
\end{array}\right] \left[\begin{array}{l}
\bm{\widetilde{v}}_{\textrm{N},\textrm{T}} \\
\,\bm{\widetilde{i}}_{\textrm{N},\textrm{R}}
\end{array}\right].
\end{aligned}
\end{equation}

\noindent By isolating the voltage $\bm{v}_{\textrm{R},\textrm{M}}$  in (\ref{eq:initial-circuit-equation-3}), we get:
\begin{equation}\label{eq:vRM}
    \bm{v}_{\textrm{R}} = \bm{K}_{\textrm{RT}}\, \bm{v}_{\textrm{G}} + \bm{K}_{\textrm{RT}}\, \bm{\widetilde{v}}_{\textrm{N},\textrm{T}} + \bm{K}_{\textrm{R}}\,\bm{\widetilde{i}}_{\textrm{N},\textrm{R}}
\end{equation}

\noindent Substituting $\bm{v}_{\textrm{R},\textrm{M}}$ from (\ref{eq:iLM-vRM}) back into (\ref{eq:vRM}), we obtain:
\begin{equation}
    \frac{R_{\textrm{in}}}{\beta}\,(\bm{i}_{\textrm{L},\textrm{M}} - \widetilde{\bm{i}}_{\textrm{LNA}})= \bm{K}_{\textrm{RT}}\, \bm{v}_{\textrm{G}} + \bm{K}_{\textrm{RT}}\, \bm{\widetilde{v}}_{\textrm{N},\textrm{T}} + \bm{K}_{\textrm{R}}\,\bm{\widetilde{i}}_{\textrm{N},\textrm{R}},
\end{equation}
which yields the desired expressions in (\ref{eq:inout})--(\ref{eq:inout2}).

\end{appendices}

\bibliographystyle{IEEEtran}
\bibliography{IEEEabrv,references}

\end{document}